\documentstyle[12pt]{article}
\newcommand{\beq}{\begin{equation}}
\newcommand{\eeq}{\end{equation}}
\newcommand{\beqn}{\begin{eqnarray}}
\newcommand{\eeqn}{\end{eqnarray}}

\def\Ga{\Gamma}
\def\de{\delta}
\def\ep{\epsilon}
\def\la{\lambda}

\def\si{\sigma}

\begin{document}
\begin{titlepage}
\begin{flushright}
      
\end{flushright}

\begin{center} 
  {\Large \bf Ultraviolet Property of Noncommutative \\Wess-Zumino-Witten Model}
\renewcommand{\thefootnote}{\fnsymbol{footnote}}
\vfill
         Ko Furuta\footnote{furuta@phys.chuo-u.ac.jp} and Takeo Inami\\
         
        Department of Physics,\\
        Chuo University\\
        Kasuga, Bunkyo-ku, Tokyo 112-8551, Japan\\
\end{center}
\vfill
\begin{abstract}
We construct noncommutative extension of the Wess-Zumino-Witten (WZW) model and study its ultraviolet property. The $\beta$-function of the $U(N)$ noncommutative WZW model resembles that of the ordinary WZW model. The $U(1)$ noncommutative model has also a nontrivial fixed point.
\end{abstract}
\vfill
\end{titlepage}

\section{Introduction}

Noncommutative field theories (NCFT) have been studied for some time in the search for a new class of field thories \cite{Con}\cite{Filk}. Four-dimensional noncommutative gauge theories arizes naturally in open string theories incorporating D-branes \cite{Con2}\cite{Cheu}\cite{Sei}, and they stimulated renewed interrest in NCFT. Perturbation analyses of a few kinds of NCFT have recentry been made \cite{Mar}\cite{Aref}, but full understanding of their quantum theory is still lacking.

In this letter we aim at constructing noncommutative extension of two-dimensional integrable field theories. Integrable field theories, e.g, sine-Gordon model, non-linear $\si$ model, Wess-Zumino-Witten model, have many important physical properties but allows computation of exact Green functions quantum mechanically. If we succeed in their noncommutative extension, the resulting noncommutative integrable field theory will provide us a concrete example of quantum NCFT. They will also give us a useful guide for understanding quantum NCFT in general.

Integrable field theories posess infinitely many conservation laws (infinite-dimensional symmetry) which generate an infinite-dimensional Lie algebra or its deformation. It is vitally important to understand how Hamiltonian and other conservation laws are constructed in NCFT.

In this letter we consider the Wess-Zumino-Witten (WZW) model \cite{Witt} and construct its noncommutative extension. The noncommutative WZW model is found to have the same infinite-dimensional symmetry as the ordinary (commutative) WZW model at their respective critical points. We are interested in the perturbative property of this theory and evaluate its $\beta$-function. We assume that there exist some regularization to deal with infrared divergences. 

In the course of writing this letter there appeared three papers \cite{Dab}, \cite{Moreno} \cite{Chu} in which the WZW model in noncommutative space is discussed. Dabrowski et al.\cite{Dab} study the noncommutative nonlinear $\si$ models. The infinite-dimensional symmetry discussed in sect.3 of our letter has also been derived by them. They have pointed out some larger symmetry. In \cite{Moreno} and \cite{Chu} the authers have derived the WZW action by integrating fermion fields in the noncommutative extension of a gauge theory.
\section{Noncommutative Extension of the WZW Model}
We take the $SU(N)$ WZW model, and define its noncommutative extension (noncommutative WZW model) by the action obtained from the WZW model by replacing product of fields by $*$-product,
\beq
I(g)=-\frac{1}{2\la^2}\int_\Sigma d^2x {\rm Tr}\left(\partial_\mu g g^{-1}\partial^\mu g g^{-1}\right)_* +\frac{k}{12\pi}\int_B {\rm Tr}\left(d\tilde g{\tilde g}^{-1}\right)^3_*~.\label{action}
\eeq
Here $\Sigma$ is the boundary of a three-dimensional manifold $B$, and $\tilde g$ is an extension of $g$ such that 
\beq
\Sigma=\partial B,\hspace{6mm}{\tilde g}(y)=g(y)~{\rm for}~y\in\partial B.
\eeq
We consider the flat manifold $\Sigma$ parametrized by ($x^0$, $x^1$), and take  the metric $g^{\mu\nu}={\rm diag}(1,-1)$. 
The $*$-product $(fg)_*$ is defined by
\beq
\left(fg\right)_*(x)\stackrel{def}{=}f*g(x)=\exp{i\xi\theta^{\mu\nu}\frac{\partial}{\partial u^{\mu}}\frac{\partial}{\partial v^{\nu}}}f(u)g(v)|_{u=v=x}~,
\eeq
where $\theta^{\mu\nu}$ is a 2nd rank antisymmetric tensor and is of the form
\beq
\left\{\theta^{\mu\nu}\right\}=\left(\begin{array}{ccc}
                    0&-1&0\\1&0&0\\0&0&0
                    \end{array}\right)~.
\eeq
The $*$-product satisfies the associativity $(fgh)_* \stackrel{def}{=}(f*(g*h))=((f*g)*h)$.
The group-valued field $\tilde g(y)$ is defined as 
\beq
\tilde g=\exp_*(i\varphi^a t_a)=\left(1+i\varphi^a t_a+\frac{1}{2!}(i\varphi^a t_a)^2+\cdots\right)_*~,
\eeq 
where $t^a$ are Hermitian matrices with the normalization ${\rm Tr}(t_at_b)=\de_{ab}$. In order that the product $\tilde g_1(y)*\tilde g_2(y)=\exp_*(i\varphi_1^at^a)*\exp_*(i\varphi_2^at^a)$ makes sense, $t^a$ have to be generators of $U(N)$ in place of $SU(N)$. Hence the action (\ref{action}) defines the noncommutative $U(N)$ WZW model.
\section{Equation of motion}
We will investigate the ultraviolet property of the model (\ref{action}). This can be made in perturbation and using the background field method. To this end we begin by parametrizing the field $g$ around the classical background $g_c$. We set $g=g_q*g_c$ and express $g_q$ in terms of quantum fluctuation $\pi^a$ as   
\beq
g_q(x)=\exp_*{\left(i\la \pi(x)\right)},~~\pi=\pi^at^a~.
\eeq 
The action is then expanded in powers of $\pi^a$ as
\beq
I(g)=I(g_c)+I_1+I_2+\cdots
\eeq
where $I_n$ represents the $n$-th order term in $\pi^a$. The first two are
\beq
I_1=-\frac{i}{\la}\int_{\Sigma}d^2x\left(g^{\mu\nu}-\frac{\la^2k}{4\pi}\ep^{\mu\nu}\right){\rm Tr}\left\{\partial_\mu g_cg_c^{-1}\partial_\nu\pi\right\}_*~,\label{ac1}
\eeq
\beq
I_2=\int_{\Sigma}d^2x{\rm Tr}\left\{\frac{1}{2}(\partial_\mu\pi)^2+\frac{1}{2}\left(g^{\mu\nu}-\frac{\la^2k}{4\pi}\ep^{\mu\nu}\right)\partial_\mu g_cg_c^{-1}[\partial_\nu\pi,\pi]\right\}_*~.\label{ac2}
\eeq
We get the equation of motion from (\ref{ac1})
\beq
\left(g^{\mu\nu}-\frac{\la^2k}{4\pi}\ep^{\mu\nu}\right){\rm Tr}\partial_\nu\left(\partial_\mu g_cg_c^{-1}\right)_*=0~.\label{motion}
\eeq

For the specific value of the coupling constant $\la$,
\beq 
\lambda^2=\frac{4\pi}{k}~,\label{crit}
\eeq 
eq.(\ref{motion}) takes a simpler form. Namely it is reduced to the left-handed (or equivalentry right-handed) current conservation law,
\beq
\partial_{+}J_{-}=0,~~J_{-}=\partial_{-}g*g^{-1}
\eeq
(or equivalentry,
\beq
\partial_{-}J_{+}=0,~~J_{+}=g^{-1}*\partial_{+}g)~.
\eeq
Here $\partial_{\pm}$ refers to the light-cone coodinates $x^{\pm}=2^{-\frac{1}{2}}(x^0\pm x^1$).
At the point (\ref{crit}), the action (\ref{action}) can be shown to be invariant under the transformations of $g$ by left and right multiplication,
\beqn
g(x)&\to& g'(x)=g_-(x^-)*g(x)~,\nonumber\\
g(x)&\to& g'(x)=g(x)*g_+(x^+)~,
\eeqn
where $g_{\pm}$ represents any element of $U(N)$ and depends only on $x^{\pm}$, respectively. The fact that $g_{\pm}$ is a function of $x^{\pm}$ means that the symmetry group is infinite-dimensional. It is a loop group if we choose $x^{\pm}$ to be a peoriodic coodinate.  

We further note the products of (an arbitrary number of) the chiral current $J_{\pm}$ are also conserved. In particular, the singlet part of quadratic terms 
\beq
T_{\pm}(x^{\pm})=:J_{\pm}^a(x^{\pm})J_{\pm}^a(x^{\pm}):
\eeq
is reminiscent of the Sugawara construction of the Virasoro algebra. This possibility needs a further study.
\section{One-loop $\beta$-function}
\begin{figure}
\begin{center}
\unitlength 0.1in
\begin{picture}(14.43,9.60)(1.30,-10.20)
%
\special{pn 8}%
\special{ar 853 540 480 480  3.2659476 6.1588303}%
%
\special{pn 8}%
\special{ar 853 540 480 480  0.1243550 3.0172377}%
%
\special{pn 8}%
\special{ar 373 540 60 60  0.0000000 6.2831853}%
%
\special{pn 8}%
\special{ar 1333 540 60 60  0.0000000 6.2831853}%
%
\special{pn 8}%
\special{pa 337 501}%
\special{pa 415 579}%
\special{fp}%
\special{pa 415 501}%
\special{pa 337 579}%
\special{fp}%
\special{pa 1372 579}%
\special{pa 1294 501}%
\special{fp}%
\special{pa 1294 579}%
\special{pa 1372 501}%
\special{fp}%
%
\special{pn 8}%
\special{pa 310 540}%
\special{pa 130 540}%
\special{da 0.070}%
\special{pa 1393 540}%
\special{pa 1573 540}%
\special{da 0.070}%
\end{picture}%
\caption{
The one-loop diagram which contributes to the UV divergence of the effective action. Dashed line reprsents the background field and solid line the propergator for $\pi$.\label{fig.1}}
\end{center}
\end{figure}
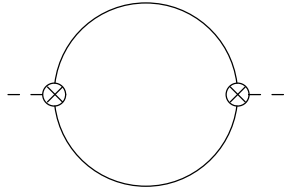

The standard background field method can be applied to the loop computation in NCFT. In this method $I_{n} (n\geq 2)$ can be regarded as the interaction term for the $\pi$ field. To evaluate one-loop terms, only $I_2$ is needed. We introduce the momentum representation of $I_2$. The interaction part (second term) of (\ref{ac2}) is then,
\beqn
I_{{\rm int}}(g_c,\pi)&=&\frac{1}{4}\frac{f^{\mu\nu}}{(2\pi)^4}\int d^2p_1 d^2p_2 d^2p_3 \de^2(p_1+p_2+p_3)G^c_\mu(p_1)\pi^a(p_2)\pi^b(p_3)\nonumber\\
            & &\times i(p_2-p_3)_{\nu}\left(f_{cab}\cos{(p_2\wedge p_3)}-id_{cab}\sin{(p_2\wedge p_3)}\right)~,\label{int}
\eeqn
where we have written 
\beqn
f^{\mu\nu}&=&g^{\mu\nu}-\frac{\la^2 k}{4\pi}\ep^{\mu\nu}\nonumber\\
p\wedge q&=&\xi\theta^{\mu\nu}p_\mu q_\nu~.
\eeqn
$G^a_\mu=\left(\partial_\mu g_c*g_c^{-1}\right)^a$ is the background. The group factors appearing in eq. (\ref{int}) are 
\beq
f_{abc}={\rm Tr}\left(t_a[t_b,t_c]\right)
,~~d_{abc}={\rm Tr}\left(t_a\{t_b,t_c\}\right)~.
\eeq
There is only one one-loop diagram that contributes to the ultraviolet (UV) divergence of the effective action $\Ga(g_c)$ (Fig. \ref{fig.1}).
The contribution of this diagram to $\Ga(g_c)$ is
\beqn
\Ga_{{\rm Fig}. 1}(g_c)&=&\frac{1}{i}\int\frac{d^2p}{(2\pi)^2}\frac{-1}{64}f^{\rho\la}f^{\mu\nu}G_{\rho}^a(p)G_{\mu}^b(-p)\int\frac{d^2k}{(2\pi)^2}\frac{(p+2k)_\la(p+2k)_\nu}{(k^2-\mu^2)\{(p+k)^2-\mu^2\}}\nonumber\\
                 & &\times\left\{(F+D)_{ab}+\cos({2p\wedge k})(F-D)_{ab}-i\sin({2p\wedge k})C_{ab}\right\}~,\label{1-loop}
\eeqn
where
\beqn
F_{ab}=f_{acd}f_{bdc}~,~~D_{ab}=d_{acd}d_{bdc}~,\nonumber\\
C_{ab}=f_{acd}d_{bdc}-f_{bcd}d_{adc}~.
\eeqn
Only the first term in the second line of (\ref{1-loop}) contributes to the UV divergence. After using the identity for $U(N)$ group
\beq
F_{ab}+D_{ab}=4N\de_{ab}~,\label{Cas.U(N)}
\eeq
we obtain the
UV divergent part of (\ref{1-loop}) as 
\beq
\Ga_{{\rm div}}^{(1-{\rm loop})}(g_c)=\frac{N\la^2}{32\pi}\left\{1-\left(\frac{\la^2k}{4\pi}\right)^2\right\}\ln{\frac{\mu^2}{\Lambda^2}}S(g_c)~,\label{1-loop div}
\eeq
where
\beq
S(g_c)=-\frac{1}{2\la^2}\int_\Sigma d^2x {\rm Tr}\left(\partial_\mu g_c g_c^{-1}\partial^\mu g_c g_c^{-1}\right)_*
\eeq
is the first term in the action (\ref{action}) for the background field. $\mu$ is the renormalization point and $\Lambda$ is the ultraviolet cutoff. We take (\ref{1-loop div})(with opposite sign) as the one-loop counterterm, then we get the $\beta$-function of $U(N)$ noncommutative WZW model as
\beq
\beta_{N.C.}=\mu\frac{d}{d\mu}\la=-\frac{N\la^3}{32\pi}\left\{1-\left(\frac{\la^2k}{4\pi}\right)^2\right\}~.\label{n.c.beta}
\eeq
As pointed out in \cite{Witt}, the second term in R.H.S of (\ref{n.c.beta}) has meaning only for large $k$.

The above result is the same as that of the ordinary $SU(N)$ WZW model \cite{Witt}. To see this, set $\xi\to0$ in (\ref{1-loop}). We then obtain the same result as (\ref{n.c.beta}), and hence
\beq
\beta=-\frac{N\la^3}{32\pi}\left\{1-\left(\frac{\la^2k}{4\pi}\right)^2\right\}~.\label{beta}
\eeq
We have used the identity for $SU(N)$ group
\beq
F_{ab}=2N\de_{ab}~
\eeq
instead of (\ref{Cas.U(N)}).
\section{Acnowledgements}
We would like to thank N. Ishibashi, M. Hayakawa and Y. Yoshida for discussions and helpful suggestions.


\end{document}